\documentclass[twocolumn,prb,superscriptaddress,preprintnumbers,amsmath,amssymb,showpacs,floatfix]{revtex4-1}

\usepackage{graphicx}
\usepackage{dcolumn}
\usepackage{bm}
\usepackage{color}

\newcommand{\ccro}{CaCu$_3$Ru$_4$O$_{12}$}
\newcommand{\lxccro}{Ca$_{1-x}$La$_x$Cu$_3$Ru$_4$O$_{12}$}
\newcommand{\lccro}{Ca$_{0.5}$La$_{0.5}$Cu$_3$Ru$_4$O$_{12}$}
\newcommand{\lcro}{LaCu$_3$Ru$_4$O$_{12}$}
\newcommand{\led}{$L_{2,3}$}

\begin{document}

\title{Correlation effects in \ccro}
\author{N.~Hollmann}
 \affiliation{Max-Planck-Institut f\"ur Chemische Physik fester Stoffe, N\"othnitzer Str. 40, 01187 Dresden, Germany}
\author{Z.~Hu}
 \affiliation{Max-Planck-Institut f\"ur Chemische Physik fester Stoffe, N\"othnitzer Str. 40, 01187 Dresden, Germany}
\author{A.~Maignan}
 \affiliation{Laboratoire CRISMAT, ENSICAEN, UMR 6508 CNRS, 6 Boulevard du Mar\'echal Juin, 14050 Caen Cedex, France}
\author{A.~G\"unther}
 \affiliation{Experimental Physics V, Center for Electronic Correlations and 
Magnetism, University of Augsburg, 86135 Augsburg, Germany}
\author{L.-Y.~Jang}
 \affiliation{National Synchrotron Radiation Research Center, Hsinchu 30076, Taiwan}
\author{A. Tanaka}
 \affiliation{Department of Quantum Matter, ADSM Hiroshima University, Higashi-Hiroshima 739-8530, Japan}
\author{H.-J. Lin}
 \affiliation{National Synchrotron Radiation Research Center, Hsinchu 30076, Taiwan}
\author{C. T. Chen}
 \affiliation{National Synchrotron Radiation Research Center, Hsinchu 30076, Taiwan}
 \author{P.~Thalmeier}
 \affiliation{Max-Planck-Institut f\"ur Chemische Physik fester Stoffe, N\"othnitzer Str. 40, 01187 Dresden, Germany}
\author{L.~H.~Tjeng}
 \affiliation{Max-Planck-Institut f\"ur Chemische Physik fester Stoffe, N\"othnitzer Str. 40, 01187 Dresden, Germany}

\date{\today}


\begin{abstract}

We have investigated the electronic structure of CaCu$_3$Ru$_4$O$_{12}$ and LaCu$_3$Ru$_4$O$_{12}$ using soft x-ray photoelectron and absorption spectroscopy together with band structure and cluster configuration interaction calculations. We found the Cu to be in a robust divalent ionic state while the Ru is more itinerant in character and stabilizes the metallic state. Substitution of Ca by La predominantly affects the Ru states. We observed strong correlation effects in the Cu $3d$ states affecting the valence band line shape considerably. Using resonant photoelectron spectroscopy at the Cu $L_3$ edge we were able to unveil the position of the Zhang-Rice singlet states in the one-electron removal spectrum of the Cu with respect to the Ru-derived metallic bands in the vicinity of the chemical potential.

\end{abstract}

\maketitle

\section{Introduction}

Transition metal oxides are known for their broad spectrum of unusual physical phenomena that originate from the correlated motion of the valence electrons of the constituent transition metal ions. Studying the interplay of two types of transition metals with different degrees of electron correlation could be an attractive way to explore for new phenomena in this field of research. The $A$-site-ordered perovskite structure $AC_3B_4$O$_{12}$ offers such a possibility. The structure is closely related to the well-known simple perovskites $AB$O$_3$ but it additionally has an ordered site $C$ occupied by a Jahn-Teller-active transition metal ion.\cite{vasiliev07a} This class of compounds has drawn considerable attention lately with the discovery of record high dielectric constants in CaCu$_3$Ti$_4$O$_{12}$\cite{subramanian00a, ramirez00a, homes01a} and giant magnetoresistance in CaCu$_3$Mn$_4$O$_{12}$,\cite{zeng99a} as well as multiferroicity.\cite{johnson12a}

\ccro\ is also an intriguing material. It exhibits unusual magnetic and transport properties.\cite{kobayashi04a, krimmel08a, krimmel09a} For example, the temperature dependence of the magnetic susceptibility resembles very much that of  CeSn$_3$, as both compounds show a broad maximum around a characteristic temperature $T^*$, which is about 200 K for \ccro. Furthermore, an enhanced quasi-particle mass has been observed from specific heat measurements. It was proposed that the underlying mechanism could be the exchange interaction of localized magnetic Cu$^{2+}$ ions with the itinerant Ru $4d$ electrons, which together constitute a Kondo lattice. This would be reminiscent of the screening of localized $4f$ moments in metallic rare earth compounds. This Kondo picture seems to be supported by photoelectron spectroscopy studies\cite{tran06a, sudayama09a} which claim that a resonance-like peak should exist at the Fermi level. If true, \ccro\ may represent the rare case of a transition-metal based Kondo system.

However, other findings oppose the Kondo scenario. Density functional calculations \cite{xiang07a} already obtain a value for the Sommerfield coefficient $\gamma$ close to the experimental one, therefore a many-body enhancement of the quasiparticle mass due to $3d$ spin fluctuations seems not required. In particular NQR/NMR investigations \cite{kato09} show that the temperature dependence of the relaxation rate is still of Korringa type above $T^*$ which speaks against the presence of $3d$ local moments necessary for the Kondo scenario. Studies on the substitution of Ca by La and Na reveal that only for \ccro\ the broad maximum in the magnetic susceptibility around $T^*$ is pronounced, while the $\gamma$ coefficients are large in all compounds, with \ccro\ having the smallest $\gamma$.\cite{tanaka09a, tanaka09b} Up to date, there is no consensus on the details of the mechanism for the enhanced Sommerfeld coefficient, the unusual magnetic behavior, and the validity of the Kondo scenario.

Here we present a soft x-ray spectroscopic study on the electronic structure of \ccro\ and \lcro. We will address the charge state of the transition metals which is related to the possible existence of local moments since this is not clear \emph{a priori} for the $A$-site-ordered perovskites. We will also discuss the effect of electron doping. A detailed examination of the valence band needs to be made by comparing the photoelectron spectroscopy results with density-functional and full-multiplet configuration-interaction calculations. The role of Cu for the states near the Fermi energy will be investigated using Cu $2p$ resonant photoelectron spectroscopy, thereby focusing in particular on the energetics of the Zhang-Rice singlets.

\section{Methods}

Polycrystalline samples of \lxccro\ have been prepared by solid-state reaction.\cite{ebbinghaus02a} The crystallographic quality of the ceramics samples was also checked by transmission electron microscopy. In particular, for the \lccro\ sample the energy-dispersive x-ray (EDX) analyses coupled to electron diffraction confirmed a good agreement between the actual and nominal cation contents. The spectroscopic experiments were carried out at the National Synchrotron Radiation Research Center (NSRRC) in Hsinchu, Taiwan. Soft x-ray absorption spectroscopy (XAS) at the Cu-\led\ and O-$K$ edges as well as soft x-ray photoelectron spectroscopy (PES) were performed at the 11A1 Dragon beam line. Clean sample surfaces were obtained by cleaving pelletized samples \emph{in situ} in an ultra-high vacuum chamber with a pressure in the low 10$^{-10}$~mbar range. The absorption spectra were recorded in total electron yield mode (TEY) and, for the O-$K$ edge, also in total fluorescence yield mode (TFY), with an energy resolution of about 0.3~eV. A Scienta SES-100 analyzer was utilized to obtain the PES spectra, with a pass energy of 100~eV. The overall energy resolution for the PES taken with photon energies in the vicinity of the Cu-\led\ edge was about 0.8~eV. The Ru-\led\ edge XAS was measured at the 16A1 tender x-ray beamline of the NSRRC. Polycrystalline powder from the samples were brushed on Kapton tape and the absorption was measured in fluorescence yield mode using a Lytle detector. All experiments were done at room temperature.

\textit{Ab initio} band structure calculations were performed by using the
full-potential augmented plane waves (FLAPW) plus the local orbital method of the {\sc Wien2k} program package.\cite{wien2k} We took the experimental
crystal structure data of \ccro\cite{subramanian02a} as input. The generalized-gradient approximation (GGA) \cite{perdew96a} was used for the exchange-correlation functional. To account for correlation effects at the Cu $3d$ within this mean-field approach we use the GGA + $U$ method \cite{anisimov93a}. We chose $U_{dd}(\mbox{Cu})=8$~eV and $J_H (\mbox{Cu})=0.8$~eV, comparable to values used for CaCuO$_2$.\cite{anisimov91a} The calculations were based on a $k$-mesh of $20\times 20\times 20$ $k$~points in the whole Brillouin zone with $R_{MT}k_{max}=7.5$. The self-consistent cycle was considered as converged for a charge difference smaller than $10^{-5}$. For the calculation of the density of states, a mesh of $50\times 50\times 50$ $k$~points was used. To model the photoelectron spectra for the Cu $3d$ we performed full-multiplet configuration-interaction calculations using the XTLS 9.0 code\cite{tanaka94a} and parameters along the lines of Eskes \emph{et al.}\cite{eskes90a}

\section{Results A: Cu and Ru valence}

\begin{figure}[t]
\includegraphics[angle=270]{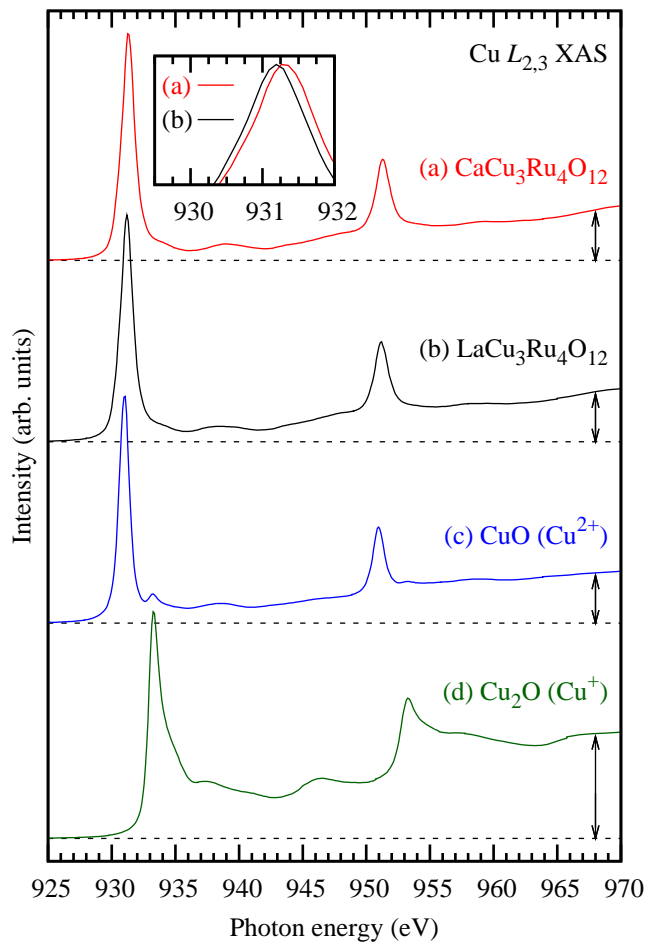}
\caption{\label{fig:cu}(Color online) Experimental Cu-\led\ x-ray absorption spectra of (a) \ccro, (b) \lcro, (c) CuO, and (d) Cu$_2$O. The spectra are normalized to the peak height and the arrows indicate the magnitude of the edge jump to the continuum states. The inset shows a close-up of the Cu-$L_3$ peak region.}
\end{figure}

In investigating the electronic structure of the \ccro\ system, first we will start with determining the valence of the Cu and Ru ions. In Fig.~\ref{fig:cu}, we show the Cu-\led\ XAS spectra of \ccro\ and \lcro, together with those of CuO and Cu$_2$O as reference for a Cu$^{2+}$ and Cu$^{+}$ compound, respectively. The spectra are dominated by the Cu 2p core-hole spin-orbit coupling, which split the spectrum roughly into two
edges, namely, the $L_3$ ($h\nu \approx$ 932 eV) and  $L_2$ ($h\nu \approx$ 952 eV) white line regions. For CuO the simple line shape of the spectrum (single line for each of the edges) is given by the dipole-allowed atomic-like Cu $2p^63d^9\to 2p^53d^{10}$ excitation and reflects the presence of a hole in the Cu $3d$ shell in the ground state of CuO.\cite{tjeng92a}

The spectrum of Cu$_2$O requires a more intricate explanation. Here the Cu $3d$ shell is formally closed in the ground state, so that the Cu $2p\to 3d$ excitation would not be possible. Nevertheless, the presence of Cu $4sp$ states and their hybridization with the O $2p$ states allows for virtual Cu $3d$ to $4sp$ excitations to occur, which in turn produce some finite amount of holes in the Cu $3d$ shell. As a result, the intensity of the Cu-\led\ edge becomes non-zero. One should notice that the ratio between the white line intensity and the edge jump to the continuum states (as indicated by the arrows in Fig.~\ref{fig:cu}) is much smaller for Cu$_2$O than for CuO. This is fully consistent with the fact that the amount of Cu $3d$ holes is obviously much smaller in Cu$_2$O than in CuO. It is also important to notice that there is an energy difference between the two spectra of about 2~eV, due to the fact that in Cu$_2$O the Cu-\led\ absorption requires a $3d\to 4sp$ virtual excitation across the gap.

We now can compare the Cu-\led\ XAS spectra of \ccro\ and \lcro\ with those of CuO and Cu$_2$O. We can observe very clearly that the energy positions of the white lines as well as the ratio of the white line intensity to the edge jump of \ccro\ and \lcro\ are very similar to those of CuO and very different from those of Cu$_2$O. We therefore can safely conclude that the Cu valence in \ccro\ and \lcro\ is 2+.  For completeness we would like to remark that there is a small energy shift of the \ccro\ and \lcro\ compared to CuO of about 220~meV. This small difference can be attributed to a difference in the amount of screening associated with the different crystal structures. A valence change of Cu between \ccro\ and \lcro\ was seen in XANES,\cite{ebbinghaus10a} but between the \ccro\ and \lcro\ Cu-\led\ spectra there is only a shift of about 0.1~eV, see the inset of Fig.~\ref{fig:cu}. 

\begin{figure}[t]
\includegraphics[angle=270]{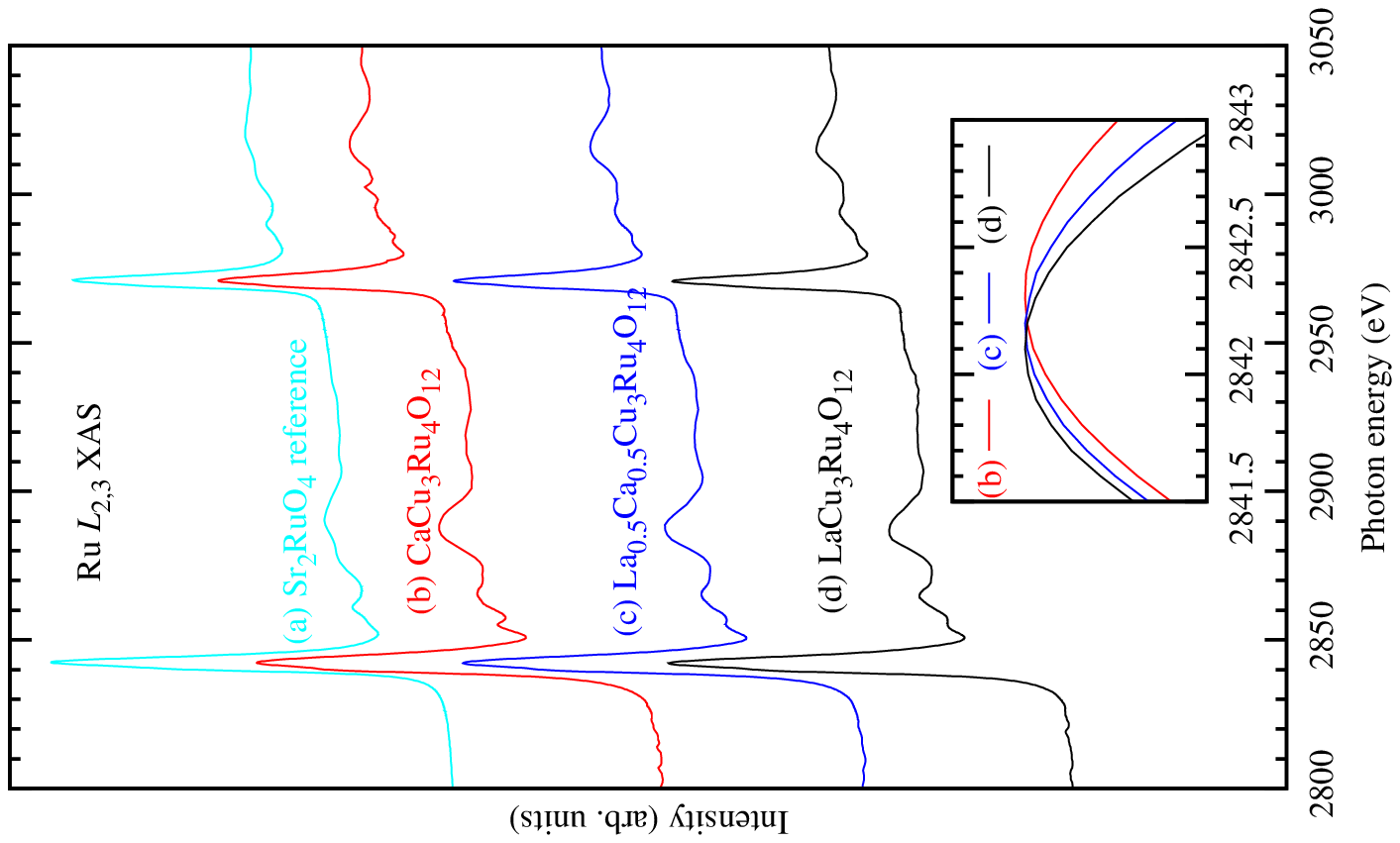}
\caption{\label{fig:ru}(Color online) Experimental Ru-\led\ x-ray absorption spectra of  Sr$_2$RuO$_4$, \ccro, \lccro, and \lcro. The inset shows a closeup of the Ru-$L_3$ peak region.}
\end{figure}

\begin{figure}[t!]
\includegraphics[angle=270]{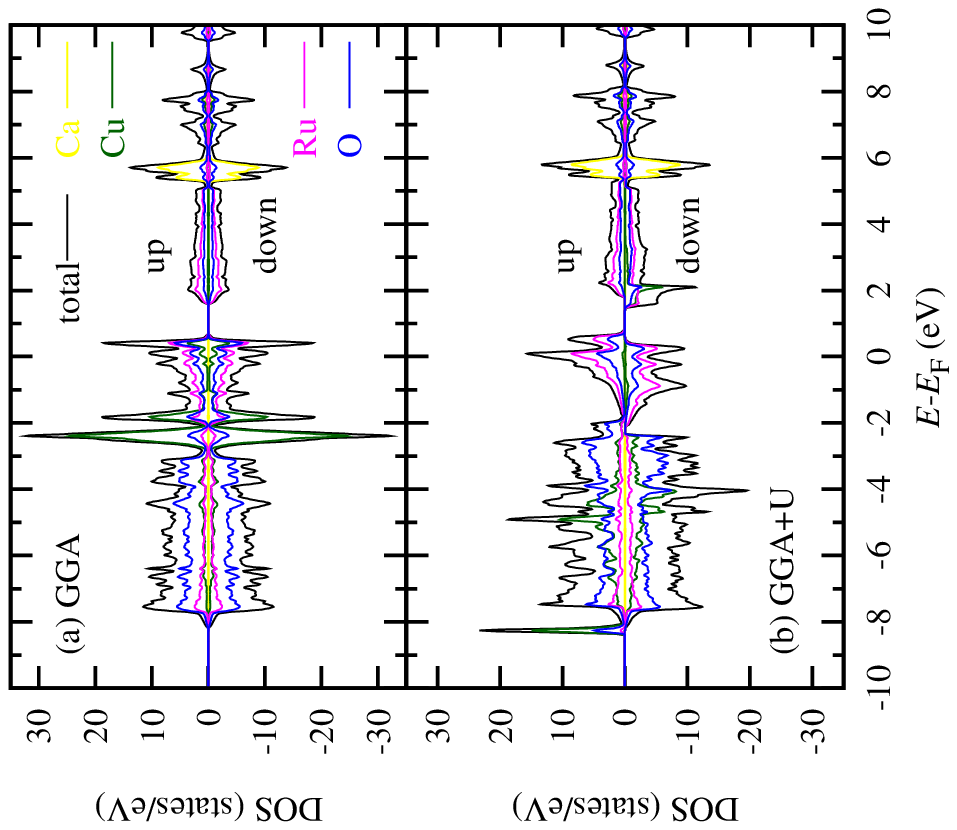}
\includegraphics[angle=270]{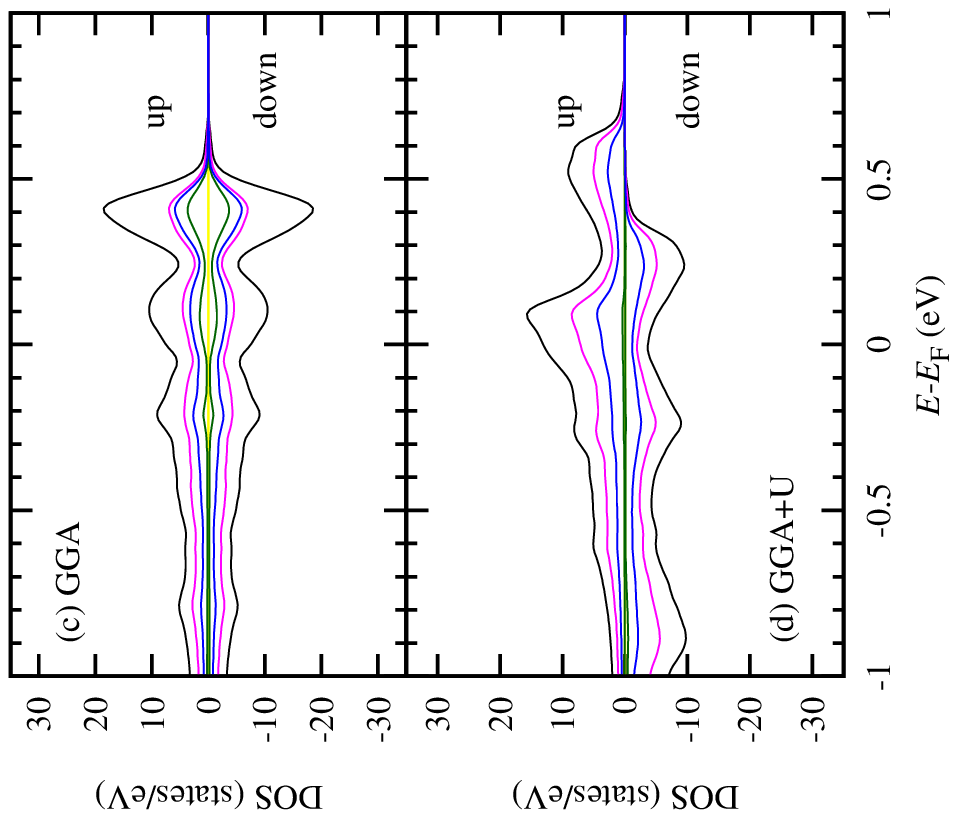}
\caption{\label{fig:dos}(Color online) Partial density of states from spin-polarized band structure calculations. (Top) (a) GGA and (b) GGA+$U$ with $U_{dd}(\mbox{Cu})=8$~eV and $J_H (\mbox{Cu})=0.8$~eV. (Bottom) Closeup of the Fermi level region for (c) GGA and (d) GGA+$U$.}
\end{figure}

To determine whether it is the Ru states that absorb the electrons upon La doping, we record the Ru-\led\ XAS for three different doping levels in Ca$_{1-x}$La$_x$Cu$_3$Ru$_4$O$_{12}$ with $x=0$, 0.5, and 1. The spectra measured in the fluorescence yield mode are shown in Fig.~\ref{fig:ru} together with the spectrum of Sr$_2$RuO$_4$ as a reference for a Ru $4d^4$ configuration with an itinerant electron character.\cite{hu02a} The energy of the $L_3$ maximum of \ccro\ is rather similar to the one of Sr$_2$RuO$_4$, indicating that the electron occupation of Ru in \ccro\ is indeed close to $4d^4$. A doping dependence for Ca$_{1-x}$La$_x$Ru$_4$O$_{12}$ is observed: for higher $x$, i.e. higher electron doping, the Ru $L_3$ absorption maximum shifts systematically towards lower energies. This indicates an increase of the number of electrons at the Ru sites: for open shells the energy position of the XAS spectrum shifts to lower energies with decreasing valence.\cite{degroot94a} At the maximum of the $L_3$ line it is seen that doping with one electron per formula unit (four Ru atoms) results in a shift of about an quarter of an eV, in agreement with the notion that a valence change of one electron results in a shift of the transition metal \led\ edge of the order of 1~eV.\cite{chang09a}

\section{Results B: Valence and Conduction Band}

Having established the Cu and Ru valences, we now focus on the band structure of the \ccro\ system.  Figure~\ref{fig:dos} (a) depicts the total, as well as the Ca, Cu, Ru, and O partial density of states (DOS) calculated using the GGA. In anticipation of the presence of correlation effects in the open Cu $3d$ shell, we also show in Fig.~\ref{fig:dos} (b) the spin-resolved total and partial DOS calculated using the GGA+$U$ with $U_{dd}(\mbox{Cu})=8$~eV and $J_H (\mbox{Cu})=0.8$~eV. A closeup of the Fermi level region is given in panels Figs.~\ref{fig:dos} (c) and \ref{fig:dos} (d), respectively.

The GGA calculations produce a non-magnetic metal for \ccro, with the Cu $3d$, Ru $4d$, and the O $2p$ states all contributing to the metallic character as can be seen more clearly from the closeup in Fig.~\ref{fig:dos} (c). The gap in the conduction band between 0.7 and 1.7 eV above the Fermi level originates from the large $e_{g}$-$t_{2g}$ crystal field splitting in the Ru $4d$. Our GGA results are similar to those of earlier band structure studies.\cite{schwingenschloegl03a, xiang07a} Interestingly, the GGA calculations show already sufficient amounts of unoccupied DOS in the Cu $3d$ as to explain the 2+ valence of the Cu ion. Upon including $U_{dd}(\mbox{Cu})$ and $J_H (\mbox{Cu})$ for the \ccro\ in the ferromagnetic state, the Cu $3d$ partial DOS is pushed away from the Fermi level region, see Fig.~\ref{fig:dos} (d), and a deep Cu $3d$ derived  spin-up state appears at the bottom of the valence band and a sharp Cu spin-down state at 2 eV above the Fermi level. Note that in this case some kind of magnetic order has to be included in the GGA+$U$ calculations in order to make $U$ active; a paramagnetic character (in contrast to a non-magnetic character), as found in the experiment, cannot be described by the DFT approach. The Cu valence is clearly 2+ with a localized moment. The Cu and Ru moments are coupled antiferromagnetically, leading to a vanishing total moment. These GGA+$U$ calculations also produce the \ccro\ to be metallic, but now with only the Ru $4d$ and O $2p$ states forming the Fermi surface. What we can learn from the GGA and GGA+$U$ calculations is that the 2+ valence of the Cu as well as the metallicity of \ccro\ are robust as these results are insensitive to the method of calculations. The number of states at the Fermi level is $N(E_F)=7.5$ states/eV and 8.5 states/eV per f.u. for GGA and GGA+$U$, respectively. The calculated values for the Sommerfeld coefficient are then 31 mJ/Cu mol K$^2$ and 35 mJ/Cu mol K$^2$, respectively, both comparable to the experimental one. This agreement shows that the enhanced mass is already correctly described in a band picture even without the inclusion of $U$.

\begin{figure}[t]
\includegraphics[angle=270]{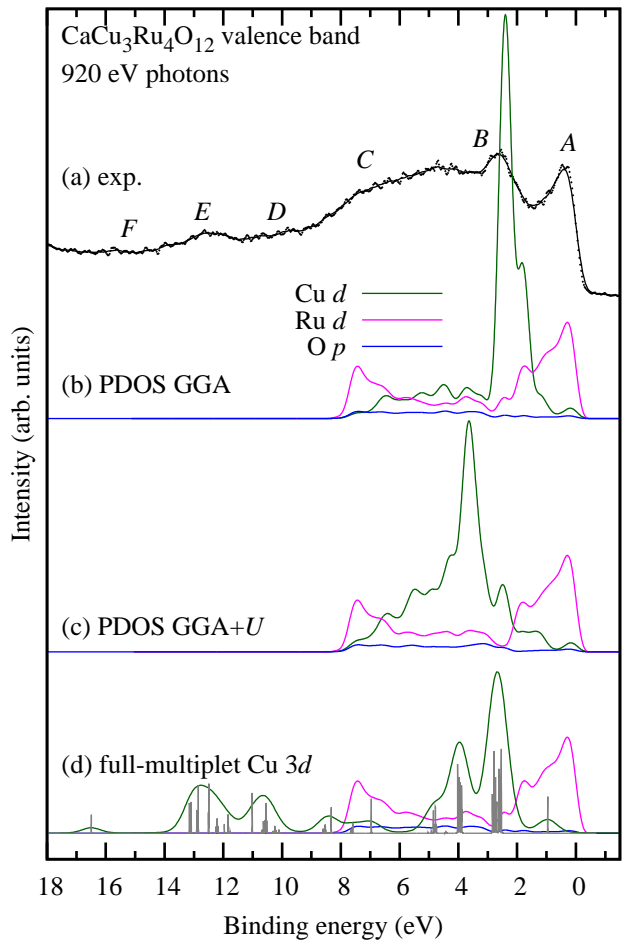}
\caption{\label{fig:pes}(Color online) Valence band spectrum of \ccro: (a) experimental photoemission results (symbols, with the thin line as a guide to the eye), (b) partial density of states calculated from GGA for Cu $d$, Ru $d$, and O $p$, weighted with the corresponding photo-ionization cross sections, (c) partial density of states from GGA+$U$, and (d) full-multiplet calculation for the Cu 3$d$ spectral weight.}
\end{figure}

To test the band structure results, we have performed PES experiments. In Fig.~\ref{fig:pes} (a) we plot the valence band spectrum of \ccro\ recorded using photons with 920~eV energy. The spectrum is comparable to those published by Tran \emph{et al.}~\cite{tran06a} and Sudayama \emph{et al.},\cite{sudayama09a} which both were taken with 1486.6~eV photons (Al $K\alpha$). Three main features labeled as $A$, $B$, and $C$ stand out. In addition, three weaker high-energy features labeled as $D$, $E$, and $F$ can be observed. We first compare the experiment with the simulation from the GGA calculations, for which we have multiplied the partial DOS by the Fermi function and broadened with a Gaussian function using $\sigma=0.35$ eV to account for the experimental resolution. We have also scaled the partial DOS of the Cu $3d$, Ru $4d$ and O $2p$ with the corresponding photo-ionization cross-sections for 920 eV photon energy.\cite{yeh85a} We can see from Fig.~\ref{fig:pes} (b) that features $A$ and $C$ can be reproduced reasonably well, with $A$ being of Ru $4d$ nature according to the calculations. However, feature $B$ which is mostly of Cu $3d$ character according to the GGA, is by far too large in the simulation. Moreover, the high-energy features $D$, $E$, and $F$ are completely missing. These discrepancies can be taken as a first evidence for the presence of electron correlation effects in the open Cu $3d$ shell.\cite{ghijsen88a,eskes90a}

As a next step, we compare the results from the GGA+$U$ calculations with the experimental valence band spectrum. Figure~\ref{fig:pes} (c) reveals that the intensity of feature $B$ in the simulation is now much too small, and that, by contrast, the 8 eV peak in the Cu $3d$ from the simulation is absent in the experiment. These disagreements between the GGA+$U$ and the experiment are not inconsistent with the correlated nature of the Cu $3d$: in fact, it is well known that while approaches using mean field $U$ corrections to band structure calculations are meant to reproduce the band gap of correlated system, one cannot expect that the calculated DOS should reproduce the photoemission spectra properly.\cite{anisimov91a,anisimov93a} Such mean field approaches are by construction not dynamic enough to handle certain excitation spectra.

We now follow an \textit{ad hoc} model. While keeping the GGA results for the Ru $4d$ and O $2p$ DOS, we will treat the Cu $3d$ spectral weight using the full-multiplet configuration-interaction calculations for a CuO$_4$ cluster in square planar symmetry as done earlier for CuO.\cite{eskes90a} The results for the Cu spectral weight are shown in Fig.~\ref{fig:pes} (d), plotted together with the results from GGA for Ru $4d$ and O $2p$. The calculated spectrum shows general agreement to the one of Eskes \emph{et al.},\cite{eskes90a} with some slight differences being caused by minor changes in the choice of parameters: we have used $U_{dd}=8.0$~eV, $\Delta=2.2$~eV, $T_{pp}=1.0$~eV, $pd\sigma =-1.23$~eV, Slater integrals reduced to 72\%\ of Hartree-Fock values. Although the agreement with the experiment is still far from perfect, we now can observe some improvements with respect to the Cu $3d$ spectral weight: peak $B$ is clearly present but not too dominating in the simulation, and the experimental high-energy features $D$, $E$, and $F$ now have their counter parts in the simulation. In fact, one can trace back these high-energy features to the multiplet structure of the Cu $3d^{8}$-line PES final states.\cite{eskes90a} We therefore can safely conclude that the open Cu $3d$ shell is the cause for strong correlation effects.

\begin{figure}[t]
\includegraphics[angle=270]{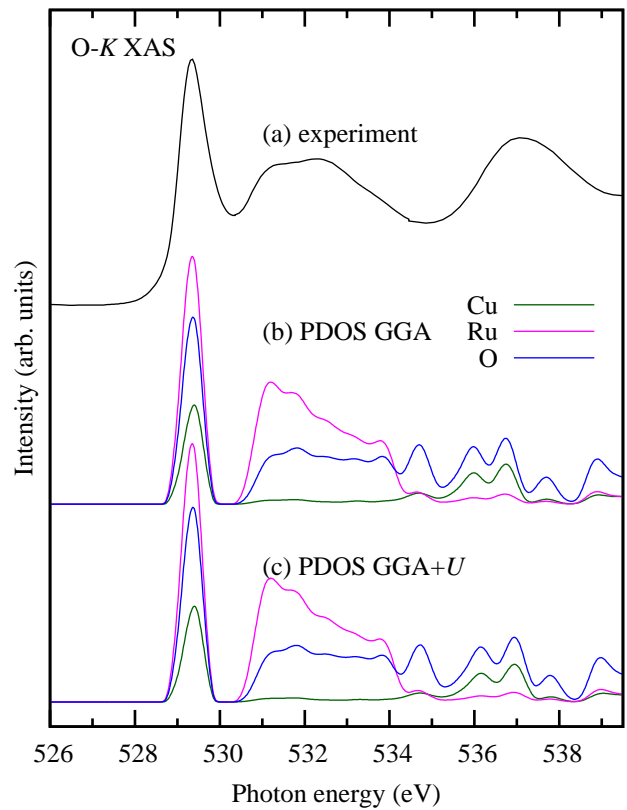}
\caption{\label{fig:o}(Color online) (a) O-$K$ x-ray absorption spectra of \ccro, together with the calculated unoccupied partial O, Ru, and Cu DOS obtained by (b) GGA and (c) GGA+$U$. The energy scale of the calculated DOS is shifted with respect to the experiment as to align the leading peak.}
\end{figure}

Information concerning the conduction band of \ccro\ can be gathered using O-$K$ XAS. This type of spectroscopy involves excitations of the type $1s^22p^5\to 1s^12p^6$ and thus provides details concerning the unoccupied O $2p$ DOS, and also of the Ru $4d$ and Cu $3d$ DOS as the formally filled $2p$ shells of the O ions are partially emptied due to hybridization with the Ru and Cu states. Figure~\ref{fig:o} shows the experimental O-$K$ spectra for \ccro. We notice that the partial unoccupied O $2p$ DOS from both the GGA and the GGA+$U$ calculations reproduce quite well the experimental spectra. This is somewhat surprising, and may be caused by the fact that the O $2p$ band width is relatively large so that electron correlation effects play a minor role. Nevertheless, correlation effects may show up indirectly at energies for which the Cu $3d$ states mix in significantly. Yet, in this particular case for which the Cu upper-Hubbard band is composed of the closed shell $3d^{10}$ configuration, the Cu derived spectral shape is free from multiplet structures and thus can be handled reasonably well by band structure methods.

It is important to note from the GGA+$U$ calculations that the leading peak is of Ru $4d$ and O $2p$ character, and that its low energy (529~eV) reflects the metallic nature of these states. It is remarkable that the GGA+$U$ calculations put the Cu upper-Hubbard band at high enough energy so that it does not fall inside the gap in the conduction band which originates from the Ru $4d$ $e_{g}$-$t_{2g}$ crystal field splitting, see Fig.~\ref{fig:dos}. As a consequence, the GGA+$U$ is able to reproduce the dip in the spectra at 530.5 eV, thereby also giving credit that the choice of $U_{dd}(\mbox{Cu})=8$~eV is reasonable and that band structure methods are capable of providing reliable estimates for the O $2p$ to Cu $3d$ charge transfer energies.\cite{eskes90a}

\section{Results C: Resonant Photoemission}

\begin{figure}[t]
\includegraphics[angle=270]{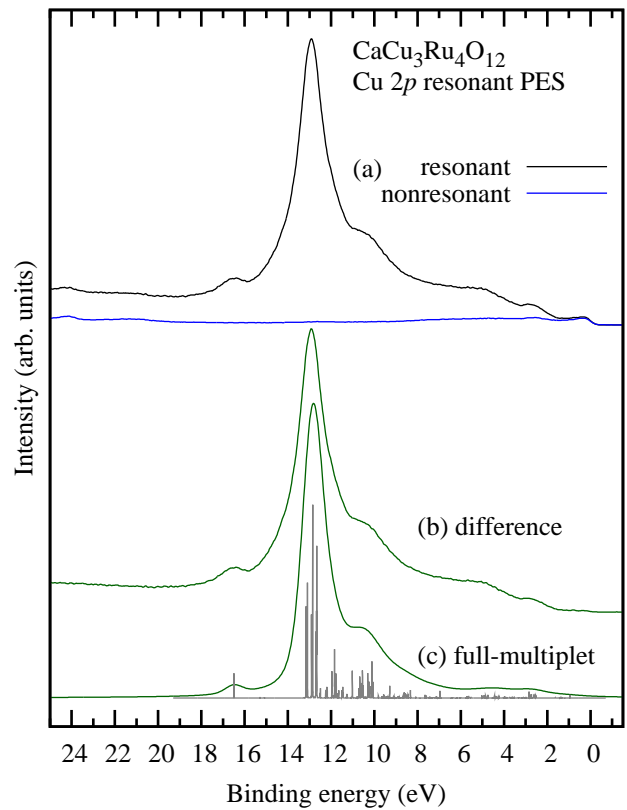}
\caption{\label{fig:rpes}(Color online) Valence band resonant photoemission of \ccro: (a) experimental valence band spectrum taken at the Cu $2p$ ($L_{3}$) resonance ($h\nu = 931.2$ eV) and away from the resonance ($h\nu = 921.2$ eV), (b) difference between the resonant and non-resonant spectra, and (c) full-multiplet simulation of the Cu $3d$ spectral weight at resonance.}
\end{figure}

To identify in more detail the contribution of the Cu $3d$ states to the electronic properties of \ccro, we performed valence band resonant PES at the Cu $2p$ ($L_{3}$) edge. Resonant PES is in particular suited to study the electronic structure of systems where correlation effects are important. For transition-metal ions like Cu, the $2p$ resonant PES leads to giant resonances that can be used to identify the energy positions of various states in the valence band. Two different photoemission channels contribute to the $2p$ resonant PES. While in normal (non-resonant) PES the excitation is given by the direct photoemission process ($3d^9+h\nu\to 3d^8+e^-$ for Cu$^{2+}$), the resonance conditions open a new channel as a source of photoelectrons. This channel can be understood as a photoabsorption process that excites $2p$ electrons into the $3d$ shell followed by the subsequent super Coster-Kronig Auger decay ($2p^63d^9+h\nu\to 2p^53d^{10}\to 2p^63d^8+e^-$ for Cu$^{2+}$).
This resonant PES technique has already been successfully applied to cuprates.\cite{tjeng91a, tjeng97a}

Figure~\ref{fig:rpes} (a) depicts the experimental valence band PES of \ccro\ taken at and off the Cu $2p$ resonance, i.e., the incident photon energy was set to the absorption maximum of the Cu-$L_3$ line ($h\nu = 931.2$~eV) and away from the resonance ($h\nu = 921.2$~eV), respectively. The on-resonance spectrum bears strong similarity to the one of CuO.\cite{tjeng91a} Huge resonance features are observed, each corresponding to different Cu $3d^8$ final states. These states are atomic-like, with the $^1G$ state at about 13~eV binding energy having the highest intensity and forming a distinct line. Within the small photon energy difference for resonant and non-resonant measurements (10~eV), the photoionization cross-sections for Cu $3d$, Ru $4d$, and O $2p$ are practically constant. The changes in the spectrum can therefore be attributed primarily to the Cu resonance process. The difference between the on- and off-resonant spectra as shown in Fig.~\ref{fig:pes} (b) then cancels the contribution of the Ru $4d$ and O $2p$ states and yields a reliable estimate for the spectral weight of Cu $3d$ in the valence band at resonance.

In Fig.~\ref{fig:rpes} (c) we also include a simulation of the Cu $3d$ spectral weight at resonance using the full-multiplet approach as described above for the nonresonant spectrum. The simulation reproduces the experiment very well. The most important parameters determining the calculated lineshape are the following: (1) the Slater integrals, which give the energy positions of the atomic-like $3d^8$ multiplet final states, and (2) the difference of $U_{dd}$ and the charge-transfer energy $\Delta$, controlling the energies of the final states with ligand-hole character ($3d^9\underline{L}$ and $3d^{10}\underline{L}^2$) relative to those of the $3d^8$.

\begin{figure}[t]
\includegraphics[angle=270]{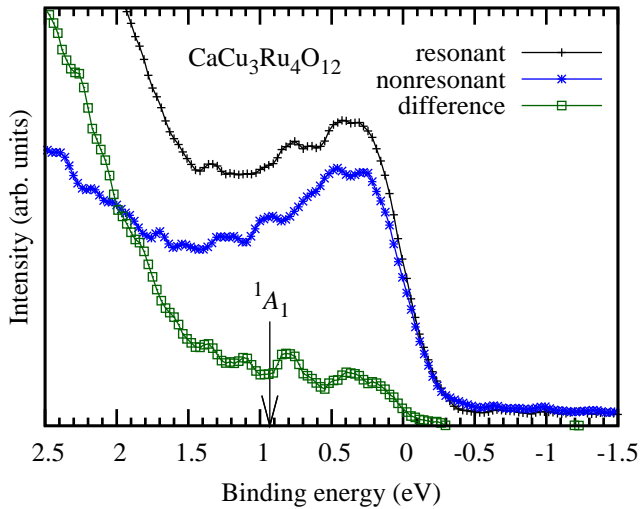}
\caption{\label{fig:ef}(Color online) Non-resonant and Cu $2p$ resonance photoelectron spectra of \ccro. The Cu $3d$ spectral weight in the vicinity of the Fermi level is deduced from the difference of resonant and non-resonant spectra.}
\end{figure}

It is now interesting to see how far the Cu $3d$ states extend up the Fermi level. Figure~\ref{fig:ef} shows a close-up of the resonant and non-resonant photoemission spectra in the region near the Fermi level of \ccro. Again, the difference of the on- and off-resonant spectra is used as an estimate for the spectral weight of Cu $3d$ in this region. We can observe that the resonant and non-resonant spectra show a Fermi cut-off. The difference spectrum on the other hand, does not have the Fermi cut-off line shape: the overlay with the Fermi function seems to show that there is a 0.1~eV distance between the edges of two curves. Although a better experimental resolution would give a more definitive answer, the present data suggest that the contribution of the Cu $3d$ states to the Fermi surface is negligible.

\section{Discussion}

For the $A$-site-ordered perovskites, the charge state of the transition metals is not clear \emph{a priori}; it was subject of discussion for a number of compounds\cite{mizokawa09b, long09b, morita10a}, including LaCu$_3$Fe$_4$O$_{12}$ \cite{long09a, *shimakawa09a} that shows evidence for valence changes with temperature. The charge state of Cu was also discussed for \ccro,\cite{ramirez04a, tran06a, krimmel09a} after valence degeneracy between Cu and Ru was proposed.\cite{subramanian02a} With XAS we show directly that Cu is in a divalent state in \ccro, in accordance with the conclusions drawn from Cu $2p$ core level XPS.\cite{tran06a, sudayama09a} Interestingly, Cu stays in this valence state upon doping, and it is the Ru which gets affected. The slight changes in intensities in core level XPS\cite{sudayama09a} are then probably the result of a different effective Ru-Cu hybridization, as it was also suggested from data on thermopower of another set of doping series.\cite{terasaki10a} 

We infer from the band structure calculations that it is the particulars of the crystal structure that form the reason for the Cu to possess a divalent state: although the Cu is embedded in a metallic background and could be expected to be well-bound and have a full $3d$ shell, the planar coordination and the resulting ligand field splitting raise the Cu $d_{x^2-y^2}$ orbital in energy. This splitting is strong enough so that the $d_{x^2-y^2}$ orbital in one spin channel is completely shifted above the Fermi energy as the GGA+$U$ calculations have shown. A $3d^9$ occupation with a localized moment is then formed. Both \ccro\ and \lcro\ possess the basic ingredients for a possible Kondo scenario: localized and correlated Cu$^{2+}$ magnetic moments embedded in a background of itinerant conduction electrons, mostly of Ru $4d$ character. It is thus remarkable that a maximum in the magnetic susceptibility is present in \ccro\ while it is absent in \lcro. 

Whether or not the presence of localized Cu$^{2+}$ moments together with the itinerant Ru electrons could lead to the formation of exotic magnetic states, comparable to the Kondo effect in rare earth compounds, depends very much on the energetics and the character of the Cu $3d$ states closest to the Fermi level. We know from earlier studies that the first ionization state of CuO and the undoped cuprates can be described as Zhang-Rice singlets.\cite{zhang88a,eskes88a,tjeng91a,tjeng97a,brookes01a} These correspond to states where the Cu ions with $3d^9$ occupation are surrounded by a hole on its oxygen ligands and they form together a bound singlet state. 
From our resonant PES experiments on \ccro\ we find indications that the Cu $3d$ spectral weight does not extend completely up to the Fermi level. There seems to be an energy distance of about 0.1~eV (or larger since the resolution of our experiment is only 0.8~eV), meaning that it will be rather unlikely that Zhang-Rice singlets can be formed by a thermal excitation of an electron from a Cu $3d^{9}$ ion to the conduction band. Our modeling of the resonant PES data using the full-multiplet calculations even suggests that these Zhang-Rice singlets are located at around 0.9~eV binding energy, quite a large value, see the mark labeled as $^{1}A_{1}$ in Fig.~\ref{fig:ef}. We thus conclude that the energetics in \ccro\ are not allowing a Kondo-like scenario. It would be very interesting if ways could be found to bring these Zhang-Rice singlets closer to the Fermi level, for example, by hole doping or by taking other $4d$ transition metal ions.

\section{Conclusions}

In summary, we have carried out x-ray absorption and photoelectron spectroscopy experiments to study the electronic structure of the \ccro\ system. We established firmly the 2+ valence state of the Cu ions. Electron doping by the substitution of Ca by La mostly affects the Ru states in what appears to be a rigid band shift. From the comparison of the experimental valence band photoemission spectra to band structure and cluster configuration interaction calculations we found that the Cu ions form local moments while the Ru are more itinerant in character and stabilize the metallic state. We observed strong correlation effects in the Cu $3d$ spectral weight affecting the valence band line shape considerably. Yet, from resonant photoemission at the Cu $2p$ ($L_3$) edge we found indications that the first ionization states of the Cu are too far from the Fermi level to allow for the formation of Zhang-Rice singlets via thermal or virtual excitations.

\section{Acknowledgements}

We wish to thank G.~A.~Sawatzky, A.~Loidl, S.~Riegg, and M.~Brando for fruitful discussions. We acknowledge support by the European Union through the ITN SOPRANO network. This work is supported by the Deutsche Forschungsgemeinschaft through FOR 1346.


\begin{thebibliography}{43}%
\makeatletter
\providecommand \@ifxundefined [1]{%
 \@ifx{#1\undefined}
}%
\providecommand \@ifnum [1]{%
 \ifnum #1\expandafter \@firstoftwo
 \else \expandafter \@secondoftwo
 \fi
}%
\providecommand \@ifx [1]{%
 \ifx #1\expandafter \@firstoftwo
 \else \expandafter \@secondoftwo
 \fi
}%
\providecommand \natexlab [1]{#1}%
\providecommand \enquote  [1]{``#1''}%
\providecommand \bibnamefont  [1]{#1}%
\providecommand \bibfnamefont [1]{#1}%
\providecommand \citenamefont [1]{#1}%
\providecommand \href@noop [0]{\@secondoftwo}%
\providecommand \href [0]{\begingroup \@sanitize@url \@href}%
\providecommand \@href[1]{\@@startlink{#1}\@@href}%
\providecommand \@@href[1]{\endgroup#1\@@endlink}%
\providecommand \@sanitize@url [0]{\catcode `\\12\catcode `\$12\catcode
  `\&12\catcode `\#12\catcode `\^12\catcode `\_12\catcode `\%12\relax}%
\providecommand \@@startlink[1]{}%
\providecommand \@@endlink[0]{}%
\providecommand \url  [0]{\begingroup\@sanitize@url \@url }%
\providecommand \@url [1]{\endgroup\@href {#1}{\urlprefix }}%
\providecommand \urlprefix  [0]{URL }%
\providecommand \Eprint [0]{\href }%
\providecommand \doibase [0]{http://dx.doi.org/}%
\providecommand \selectlanguage [0]{\@gobble}%
\providecommand \bibinfo  [0]{\@secondoftwo}%
\providecommand \bibfield  [0]{\@secondoftwo}%
\providecommand \translation [1]{[#1]}%
\providecommand \BibitemOpen [0]{}%
\providecommand \bibitemStop [0]{}%
\providecommand \bibitemNoStop [0]{.\EOS\space}%
\providecommand \EOS [0]{\spacefactor3000\relax}%
\providecommand \BibitemShut  [1]{\csname bibitem#1\endcsname}%
\let\auto@bib@innerbib\@empty
\bibitem [{\citenamefont {Vasil'ev}\ and\ \citenamefont
  {Volkova}(2007)}]{vasiliev07a}%
  \BibitemOpen
  \bibfield  {author} {\bibinfo {author} {\bibfnamefont {A.~N.}\ \bibnamefont
  {Vasil'ev}}\ and\ \bibinfo {author} {\bibfnamefont {O.~S.}\ \bibnamefont
  {Volkova}},\ }\href {\doibase 10.1063/1.2747047} {\bibfield  {journal}
  {\bibinfo  {journal} {Low Temp. Phys.}\ }\textbf {\bibinfo {volume} {33}},\
  \bibinfo {pages} {895} (\bibinfo {year} {2007})}\BibitemShut {NoStop}%
\bibitem [{\citenamefont {Subramanian}\ \emph {et~al.}(2000)\citenamefont
  {Subramanian}, \citenamefont {Li}, \citenamefont {Duan}, \citenamefont
  {Reisner},\ and\ \citenamefont {Sleight}}]{subramanian00a}%
  \BibitemOpen
  \bibfield  {author} {\bibinfo {author} {\bibfnamefont {M.}~\bibnamefont
  {Subramanian}}, \bibinfo {author} {\bibfnamefont {D.}~\bibnamefont {Li}},
  \bibinfo {author} {\bibfnamefont {N.}~\bibnamefont {Duan}}, \bibinfo {author}
  {\bibfnamefont {B.}~\bibnamefont {Reisner}}, \ and\ \bibinfo {author}
  {\bibfnamefont {A.}~\bibnamefont {Sleight}},\ }\href {\doibase
  10.1006/jssc.2000.8703} {\bibfield  {journal} {\bibinfo  {journal} {J. Solid
  State Chem.}\ }\textbf {\bibinfo {volume} {151}},\ \bibinfo {pages} {323 }
  (\bibinfo {year} {2000})}\BibitemShut {NoStop}%
\bibitem [{\citenamefont {Ramirez}\ \emph {et~al.}(2000)\citenamefont
  {Ramirez}, \citenamefont {Subramanian}, \citenamefont {Gardel}, \citenamefont
  {Blumberg}, \citenamefont {Li}, \citenamefont {Vogt},\ and\ \citenamefont
  {Shapiro}}]{ramirez00a}%
  \BibitemOpen
  \bibfield  {author} {\bibinfo {author} {\bibfnamefont {A.}~\bibnamefont
  {Ramirez}}, \bibinfo {author} {\bibfnamefont {M.}~\bibnamefont
  {Subramanian}}, \bibinfo {author} {\bibfnamefont {M.}~\bibnamefont {Gardel}},
  \bibinfo {author} {\bibfnamefont {G.}~\bibnamefont {Blumberg}}, \bibinfo
  {author} {\bibfnamefont {D.}~\bibnamefont {Li}}, \bibinfo {author}
  {\bibfnamefont {T.}~\bibnamefont {Vogt}}, \ and\ \bibinfo {author}
  {\bibfnamefont {S.}~\bibnamefont {Shapiro}},\ }\href {\doibase
  10.1016/S0038-1098(00)00182-4} {\bibfield  {journal} {\bibinfo  {journal}
  {Solid State Commun.}\ }\textbf {\bibinfo {volume} {115}},\ \bibinfo {pages}
  {217 } (\bibinfo {year} {2000})}\BibitemShut {NoStop}%
\bibitem [{\citenamefont {Homes}\ \emph {et~al.}(2001)\citenamefont {Homes},
  \citenamefont {Vogt}, \citenamefont {Shapiro}, \citenamefont {Wakimoto},\
  and\ \citenamefont {Ramirez}}]{homes01a}%
  \BibitemOpen
  \bibfield  {author} {\bibinfo {author} {\bibfnamefont {C.~C.}\ \bibnamefont
  {Homes}}, \bibinfo {author} {\bibfnamefont {T.}~\bibnamefont {Vogt}},
  \bibinfo {author} {\bibfnamefont {S.~M.}\ \bibnamefont {Shapiro}}, \bibinfo
  {author} {\bibfnamefont {S.}~\bibnamefont {Wakimoto}}, \ and\ \bibinfo
  {author} {\bibfnamefont {A.~P.}\ \bibnamefont {Ramirez}},\ }\href {\doibase
  10.1126/science.1061655} {\bibfield  {journal} {\bibinfo  {journal}
  {Science}\ }\textbf {\bibinfo {volume} {293}},\ \bibinfo {pages} {673}
  (\bibinfo {year} {2001})}\BibitemShut {NoStop}%
\bibitem [{\citenamefont {Zeng}\ \emph {et~al.}(1999)\citenamefont {Zeng},
  \citenamefont {Greenblatt}, \citenamefont {Subramanian},\ and\ \citenamefont
  {Croft}}]{zeng99a}%
  \BibitemOpen
  \bibfield  {author} {\bibinfo {author} {\bibfnamefont {Z.}~\bibnamefont
  {Zeng}}, \bibinfo {author} {\bibfnamefont {M.}~\bibnamefont {Greenblatt}},
  \bibinfo {author} {\bibfnamefont {M.~A.}\ \bibnamefont {Subramanian}}, \ and\
  \bibinfo {author} {\bibfnamefont {M.}~\bibnamefont {Croft}},\ }\href
  {\doibase 10.1103/PhysRevLett.82.3164} {\bibfield  {journal} {\bibinfo
  {journal} {Phys. Rev. Lett.}\ }\textbf {\bibinfo {volume} {82}},\ \bibinfo
  {pages} {3164} (\bibinfo {year} {1999})}\BibitemShut {NoStop}%
\bibitem [{\citenamefont {Johnson}\ \emph {et~al.}(2012)\citenamefont
  {Johnson}, \citenamefont {Chapon}, \citenamefont {Khalyavin}, \citenamefont
  {Manuel}, \citenamefont {Radaelli},\ and\ \citenamefont
  {Martin}}]{johnson12a}%
  \BibitemOpen
  \bibfield  {author} {\bibinfo {author} {\bibfnamefont {R.~D.}\ \bibnamefont
  {Johnson}}, \bibinfo {author} {\bibfnamefont {L.~C.}\ \bibnamefont {Chapon}},
  \bibinfo {author} {\bibfnamefont {D.~D.}\ \bibnamefont {Khalyavin}}, \bibinfo
  {author} {\bibfnamefont {P.}~\bibnamefont {Manuel}}, \bibinfo {author}
  {\bibfnamefont {P.~G.}\ \bibnamefont {Radaelli}}, \ and\ \bibinfo {author}
  {\bibfnamefont {C.}~\bibnamefont {Martin}},\ }\href {\doibase
  10.1103/PhysRevLett.108.067201} {\bibfield  {journal} {\bibinfo  {journal}
  {Phys. Rev. Lett.}\ }\textbf {\bibinfo {volume} {108}},\ \bibinfo {pages}
  {067201} (\bibinfo {year} {2012})}\BibitemShut {NoStop}%
\bibitem [{\citenamefont {Kobayashi}\ \emph {et~al.}(2004)\citenamefont
  {Kobayashi}, \citenamefont {Terasaki}, \citenamefont {Takeya}, \citenamefont
  {Tsukada},\ and\ \citenamefont {Ando}}]{kobayashi04a}%
  \BibitemOpen
  \bibfield  {author} {\bibinfo {author} {\bibfnamefont {W.}~\bibnamefont
  {Kobayashi}}, \bibinfo {author} {\bibfnamefont {I.}~\bibnamefont {Terasaki}},
  \bibinfo {author} {\bibfnamefont {J.}~\bibnamefont {Takeya}}, \bibinfo
  {author} {\bibfnamefont {I.}~\bibnamefont {Tsukada}}, \ and\ \bibinfo
  {author} {\bibfnamefont {Y.}~\bibnamefont {Ando}},\ }\href {\doibase
  10.1143/JPSJ.73.2373} {\bibfield  {journal} {\bibinfo  {journal} {J. Phys.
  Soc. Jpn.}\ }\textbf {\bibinfo {volume} {73}},\ \bibinfo {pages} {2373}
  (\bibinfo {year} {2004})}\BibitemShut {NoStop}%
\bibitem [{\citenamefont {Krimmel}\ \emph {et~al.}(2008)\citenamefont
  {Krimmel}, \citenamefont {G\"unther}, \citenamefont {Kraetschmer},
  \citenamefont {Dekinger}, \citenamefont {B\"uttgen}, \citenamefont {Loidl},
  \citenamefont {Ebbinghaus}, \citenamefont {Scheidt},\ and\ \citenamefont
  {Scherer}}]{krimmel08a}%
  \BibitemOpen
  \bibfield  {author} {\bibinfo {author} {\bibfnamefont {A.}~\bibnamefont
  {Krimmel}}, \bibinfo {author} {\bibfnamefont {A.}~\bibnamefont {G\"unther}},
  \bibinfo {author} {\bibfnamefont {W.}~\bibnamefont {Kraetschmer}}, \bibinfo
  {author} {\bibfnamefont {H.}~\bibnamefont {Dekinger}}, \bibinfo {author}
  {\bibfnamefont {N.}~\bibnamefont {B\"uttgen}}, \bibinfo {author}
  {\bibfnamefont {A.}~\bibnamefont {Loidl}}, \bibinfo {author} {\bibfnamefont
  {S.~G.}\ \bibnamefont {Ebbinghaus}}, \bibinfo {author} {\bibfnamefont
  {E.-W.}\ \bibnamefont {Scheidt}}, \ and\ \bibinfo {author} {\bibfnamefont
  {W.}~\bibnamefont {Scherer}},\ }\href {\doibase 10.1103/PhysRevB.78.165126}
  {\bibfield  {journal} {\bibinfo  {journal} {Phys. Rev. B}\ }\textbf {\bibinfo
  {volume} {78}},\ \bibinfo {pages} {165126} (\bibinfo {year}
  {2008})}\BibitemShut {NoStop}%
\bibitem [{\citenamefont {Krimmel}\ \emph {et~al.}(2009)\citenamefont
  {Krimmel}, \citenamefont {G\"unther}, \citenamefont {Kraetschmer},
  \citenamefont {Dekinger}, \citenamefont {B\"uttgen}, \citenamefont {Eyert},
  \citenamefont {Loidl}, \citenamefont {Sheptyakov}, \citenamefont {Scheidt},\
  and\ \citenamefont {Scherer}}]{krimmel09a}%
  \BibitemOpen
  \bibfield  {author} {\bibinfo {author} {\bibfnamefont {A.}~\bibnamefont
  {Krimmel}}, \bibinfo {author} {\bibfnamefont {A.}~\bibnamefont {G\"unther}},
  \bibinfo {author} {\bibfnamefont {W.}~\bibnamefont {Kraetschmer}}, \bibinfo
  {author} {\bibfnamefont {H.}~\bibnamefont {Dekinger}}, \bibinfo {author}
  {\bibfnamefont {N.}~\bibnamefont {B\"uttgen}}, \bibinfo {author}
  {\bibfnamefont {V.}~\bibnamefont {Eyert}}, \bibinfo {author} {\bibfnamefont
  {A.}~\bibnamefont {Loidl}}, \bibinfo {author} {\bibfnamefont {D.~V.}\
  \bibnamefont {Sheptyakov}}, \bibinfo {author} {\bibfnamefont {E.-W.}\
  \bibnamefont {Scheidt}}, \ and\ \bibinfo {author} {\bibfnamefont
  {W.}~\bibnamefont {Scherer}},\ }\href {\doibase 10.1103/PhysRevB.80.121101}
  {\bibfield  {journal} {\bibinfo  {journal} {Phys. Rev. B}\ }\textbf {\bibinfo
  {volume} {80}},\ \bibinfo {pages} {121101} (\bibinfo {year}
  {2009})}\BibitemShut {NoStop}%
\bibitem [{\citenamefont {Tran}\ \emph {et~al.}(2006)\citenamefont {Tran},
  \citenamefont {Takubo}, \citenamefont {Mizokawa}, \citenamefont {Kobayashi},\
  and\ \citenamefont {Terasaki}}]{tran06a}%
  \BibitemOpen
  \bibfield  {author} {\bibinfo {author} {\bibfnamefont {T.~T.}\ \bibnamefont
  {Tran}}, \bibinfo {author} {\bibfnamefont {K.}~\bibnamefont {Takubo}},
  \bibinfo {author} {\bibfnamefont {T.}~\bibnamefont {Mizokawa}}, \bibinfo
  {author} {\bibfnamefont {W.}~\bibnamefont {Kobayashi}}, \ and\ \bibinfo
  {author} {\bibfnamefont {I.}~\bibnamefont {Terasaki}},\ }\href {\doibase
  10.1103/PhysRevB.73.193105} {\bibfield  {journal} {\bibinfo  {journal} {Phys.
  Rev. B}\ }\textbf {\bibinfo {volume} {73}},\ \bibinfo {pages} {193105}
  (\bibinfo {year} {2006})}\BibitemShut {NoStop}%
\bibitem [{\citenamefont {Sudayama}\ \emph {et~al.}(2009)\citenamefont
  {Sudayama}, \citenamefont {Wakisaka}, \citenamefont {Takubo}, \citenamefont
  {Mizokawa}, \citenamefont {Kobayashi}, \citenamefont {Terasaki},
  \citenamefont {Tanaka}, \citenamefont {Maeno}, \citenamefont {Arita},
  \citenamefont {Namatame},\ and\ \citenamefont {Taniguchi}}]{sudayama09a}%
  \BibitemOpen
  \bibfield  {author} {\bibinfo {author} {\bibfnamefont {T.}~\bibnamefont
  {Sudayama}}, \bibinfo {author} {\bibfnamefont {Y.}~\bibnamefont {Wakisaka}},
  \bibinfo {author} {\bibfnamefont {K.}~\bibnamefont {Takubo}}, \bibinfo
  {author} {\bibfnamefont {T.}~\bibnamefont {Mizokawa}}, \bibinfo {author}
  {\bibfnamefont {W.}~\bibnamefont {Kobayashi}}, \bibinfo {author}
  {\bibfnamefont {I.}~\bibnamefont {Terasaki}}, \bibinfo {author}
  {\bibfnamefont {S.}~\bibnamefont {Tanaka}}, \bibinfo {author} {\bibfnamefont
  {Y.}~\bibnamefont {Maeno}}, \bibinfo {author} {\bibfnamefont
  {M.}~\bibnamefont {Arita}}, \bibinfo {author} {\bibfnamefont
  {H.}~\bibnamefont {Namatame}}, \ and\ \bibinfo {author} {\bibfnamefont
  {M.}~\bibnamefont {Taniguchi}},\ }\href {\doibase 10.1103/PhysRevB.80.075113}
  {\bibfield  {journal} {\bibinfo  {journal} {Phys. Rev. B}\ }\textbf {\bibinfo
  {volume} {80}},\ \bibinfo {pages} {075113} (\bibinfo {year}
  {2009})}\BibitemShut {NoStop}%
\bibitem [{\citenamefont {Xiang}\ \emph {et~al.}(2007)\citenamefont {Xiang},
  \citenamefont {Liu}, \citenamefont {Zhao}, \citenamefont {Meng},\ and\
  \citenamefont {Wu}}]{xiang07a}%
  \BibitemOpen
  \bibfield  {author} {\bibinfo {author} {\bibfnamefont {H.}~\bibnamefont
  {Xiang}}, \bibinfo {author} {\bibfnamefont {X.}~\bibnamefont {Liu}}, \bibinfo
  {author} {\bibfnamefont {E.}~\bibnamefont {Zhao}}, \bibinfo {author}
  {\bibfnamefont {J.}~\bibnamefont {Meng}}, \ and\ \bibinfo {author}
  {\bibfnamefont {Z.}~\bibnamefont {Wu}},\ }\href {\doibase
  10.1103/PhysRevB.76.155103} {\bibfield  {journal} {\bibinfo  {journal} {Phys.
  Rev. B}\ }\textbf {\bibinfo {volume} {76}},\ \bibinfo {pages} {155103}
  (\bibinfo {year} {2007})}\BibitemShut {NoStop}%
\bibitem [{\citenamefont {Kato}\ \emph {et~al.}(2009)\citenamefont {Kato},
  \citenamefont {Tsuruta},\ and\ \citenamefont {Matsumura}}]{kato09}%
  \BibitemOpen
  \bibfield  {author} {\bibinfo {author} {\bibfnamefont {H.}~\bibnamefont
  {Kato}}, \bibinfo {author} {\bibfnamefont {T.}~\bibnamefont {Tsuruta}}, \
  and\ \bibinfo {author} {\bibfnamefont {M.}~\bibnamefont {Matsumura}},\
  }\href@noop {} {\bibfield  {journal} {\bibinfo  {journal} {J. Phys. Soc.
  Jpn.}\ }\textbf {\bibinfo {volume} {78}},\ \bibinfo {pages} {054707}
  (\bibinfo {year} {2009})}\BibitemShut {NoStop}%
\bibitem [{\citenamefont {Tanaka}\ \emph
  {et~al.}(2009{\natexlab{a}})\citenamefont {Tanaka}, \citenamefont {Takatsu},
  \citenamefont {Yonezawa},\ and\ \citenamefont {Maeno}}]{tanaka09a}%
  \BibitemOpen
  \bibfield  {author} {\bibinfo {author} {\bibfnamefont {S.}~\bibnamefont
  {Tanaka}}, \bibinfo {author} {\bibfnamefont {H.}~\bibnamefont {Takatsu}},
  \bibinfo {author} {\bibfnamefont {S.}~\bibnamefont {Yonezawa}}, \ and\
  \bibinfo {author} {\bibfnamefont {Y.}~\bibnamefont {Maeno}},\ }\href
  {\doibase 10.1103/PhysRevB.80.035113} {\bibfield  {journal} {\bibinfo
  {journal} {Phys. Rev. B}\ }\textbf {\bibinfo {volume} {80}},\ \bibinfo
  {pages} {035113} (\bibinfo {year} {2009}{\natexlab{a}})}\BibitemShut
  {NoStop}%
\bibitem [{\citenamefont {Tanaka}\ \emph
  {et~al.}(2009{\natexlab{b}})\citenamefont {Tanaka}, \citenamefont {Shimazui},
  \citenamefont {Takatsu}, \citenamefont {Yonezawa},\ and\ \citenamefont
  {Maeno}}]{tanaka09b}%
  \BibitemOpen
  \bibfield  {author} {\bibinfo {author} {\bibfnamefont {S.}~\bibnamefont
  {Tanaka}}, \bibinfo {author} {\bibfnamefont {N.}~\bibnamefont {Shimazui}},
  \bibinfo {author} {\bibfnamefont {H.}~\bibnamefont {Takatsu}}, \bibinfo
  {author} {\bibfnamefont {S.}~\bibnamefont {Yonezawa}}, \ and\ \bibinfo
  {author} {\bibfnamefont {Y.}~\bibnamefont {Maeno}},\ }\href {\doibase
  10.1143/JPSJ.78.024706} {\bibfield  {journal} {\bibinfo  {journal} {J. Phys.
  Soc. Jpn.}\ }\textbf {\bibinfo {volume} {78}},\ \bibinfo {pages} {024706}
  (\bibinfo {year} {2009}{\natexlab{b}})}\BibitemShut {NoStop}%
\bibitem [{\citenamefont {Ebbinghaus}\ \emph {et~al.}(2002)\citenamefont
  {Ebbinghaus}, \citenamefont {Weidenkaff},\ and\ \citenamefont
  {Cava}}]{ebbinghaus02a}%
  \BibitemOpen
  \bibfield  {author} {\bibinfo {author} {\bibfnamefont {S.}~\bibnamefont
  {Ebbinghaus}}, \bibinfo {author} {\bibfnamefont {A.}~\bibnamefont
  {Weidenkaff}}, \ and\ \bibinfo {author} {\bibfnamefont {R.}~\bibnamefont
  {Cava}},\ }\href {\doibase 10.1006/jssc.2002.9634} {\bibfield  {journal}
  {\bibinfo  {journal} {J. Solid State Chem.}\ }\textbf {\bibinfo {volume}
  {167}},\ \bibinfo {pages} {126 } (\bibinfo {year} {2002})}\BibitemShut
  {NoStop}%
\bibitem [{\citenamefont {Blaha}\ \emph {et~al.}(2001)\citenamefont {Blaha},
  \citenamefont {Schwarz}, \citenamefont {Madsen}, \citenamefont {Kvasnicka},\
  and\ \citenamefont {Luitz}}]{wien2k}%
  \BibitemOpen
  \bibfield  {author} {\bibinfo {author} {\bibfnamefont {P.}~\bibnamefont
  {Blaha}}, \bibinfo {author} {\bibfnamefont {K.}~\bibnamefont {Schwarz}},
  \bibinfo {author} {\bibfnamefont {G.}~\bibnamefont {Madsen}}, \bibinfo
  {author} {\bibfnamefont {D.}~\bibnamefont {Kvasnicka}}, \ and\ \bibinfo
  {author} {\bibfnamefont {J.}~\bibnamefont {Luitz}},\ }\href@noop {} {\emph
  {\bibinfo {title} {Wien2k, An Augmented Plane Wave + Local Orbitals Program
  for Calculating Crystal Properties}}}\ (\bibinfo  {publisher} {K. Schwarz,
  Tech. Universit\"at Wien},\ \bibinfo {address} {Austria},\ \bibinfo {year}
  {2001})\BibitemShut {NoStop}%
\bibitem [{\citenamefont {Subramanian}\ and\ \citenamefont
  {Sleight}(2002)}]{subramanian02a}%
  \BibitemOpen
  \bibfield  {author} {\bibinfo {author} {\bibfnamefont {M.}~\bibnamefont
  {Subramanian}}\ and\ \bibinfo {author} {\bibfnamefont {A.}~\bibnamefont
  {Sleight}},\ }\href {\doibase 10.1016/S1293-2558(01)01262-6} {\bibfield
  {journal} {\bibinfo  {journal} {Solid State Sci.}\ }\textbf {\bibinfo
  {volume} {4}},\ \bibinfo {pages} {347 } (\bibinfo {year} {2002})}\BibitemShut
  {NoStop}%
\bibitem [{\citenamefont {Perdew}\ \emph {et~al.}(1996)\citenamefont {Perdew},
  \citenamefont {Burke},\ and\ \citenamefont {Ernzerhof}}]{perdew96a}%
  \BibitemOpen
  \bibfield  {author} {\bibinfo {author} {\bibfnamefont {J.~P.}\ \bibnamefont
  {Perdew}}, \bibinfo {author} {\bibfnamefont {K.}~\bibnamefont {Burke}}, \
  and\ \bibinfo {author} {\bibfnamefont {M.}~\bibnamefont {Ernzerhof}},\ }\href
  {\doibase 10.1103/PhysRevLett.77.3865} {\bibfield  {journal} {\bibinfo
  {journal} {Phys. Rev. Lett.}\ }\textbf {\bibinfo {volume} {77}},\ \bibinfo
  {pages} {3865} (\bibinfo {year} {1996})}\BibitemShut {NoStop}%
\bibitem [{\citenamefont {Anisimov}\ \emph {et~al.}(1993)\citenamefont
  {Anisimov}, \citenamefont {Solovyev}, \citenamefont {Korotin}, \citenamefont
  {Czy\ifmmode~\dot{z}\else \.{z}\fi{}yk},\ and\ \citenamefont
  {Sawatzky}}]{anisimov93a}%
  \BibitemOpen
  \bibfield  {author} {\bibinfo {author} {\bibfnamefont {V.~I.}\ \bibnamefont
  {Anisimov}}, \bibinfo {author} {\bibfnamefont {I.~V.}\ \bibnamefont
  {Solovyev}}, \bibinfo {author} {\bibfnamefont {M.~A.}\ \bibnamefont
  {Korotin}}, \bibinfo {author} {\bibfnamefont {M.~T.}\ \bibnamefont
  {Czy\ifmmode~\dot{z}\else \.{z}\fi{}yk}}, \ and\ \bibinfo {author}
  {\bibfnamefont {G.~A.}\ \bibnamefont {Sawatzky}},\ }\href {\doibase
  10.1103/PhysRevB.48.16929} {\bibfield  {journal} {\bibinfo  {journal} {Phys.
  Rev. B}\ }\textbf {\bibinfo {volume} {48}},\ \bibinfo {pages} {16929}
  (\bibinfo {year} {1993})}\BibitemShut {NoStop}%
\bibitem [{\citenamefont {Anisimov}\ \emph {et~al.}(1991)\citenamefont
  {Anisimov}, \citenamefont {Zaanen},\ and\ \citenamefont
  {Andersen}}]{anisimov91a}%
  \BibitemOpen
  \bibfield  {author} {\bibinfo {author} {\bibfnamefont {V.~I.}\ \bibnamefont
  {Anisimov}}, \bibinfo {author} {\bibfnamefont {J.}~\bibnamefont {Zaanen}}, \
  and\ \bibinfo {author} {\bibfnamefont {O.~K.}\ \bibnamefont {Andersen}},\
  }\href {\doibase 10.1103/PhysRevB.44.943} {\bibfield  {journal} {\bibinfo
  {journal} {Phys. Rev. B}\ }\textbf {\bibinfo {volume} {44}},\ \bibinfo
  {pages} {943} (\bibinfo {year} {1991})}\BibitemShut {NoStop}%
\bibitem [{\citenamefont {Tanaka}\ and\ \citenamefont {Jo}(1994)}]{tanaka94a}%
  \BibitemOpen
  \bibfield  {author} {\bibinfo {author} {\bibfnamefont {A.}~\bibnamefont
  {Tanaka}}\ and\ \bibinfo {author} {\bibfnamefont {T.}~\bibnamefont {Jo}},\
  }\href {\doibase 10.1143/JPSJ.63.2788} {\bibfield  {journal} {\bibinfo
  {journal} {J. Phys. Soc. Jpn.}\ }\textbf {\bibinfo {volume} {63}},\ \bibinfo
  {pages} {2788} (\bibinfo {year} {1994})}\BibitemShut {NoStop}%
\bibitem [{\citenamefont {Eskes}\ \emph {et~al.}(1990)\citenamefont {Eskes},
  \citenamefont {Tjeng},\ and\ \citenamefont {Sawatzky}}]{eskes90a}%
  \BibitemOpen
  \bibfield  {author} {\bibinfo {author} {\bibfnamefont {H.}~\bibnamefont
  {Eskes}}, \bibinfo {author} {\bibfnamefont {L.~H.}\ \bibnamefont {Tjeng}}, \
  and\ \bibinfo {author} {\bibfnamefont {G.~A.}\ \bibnamefont {Sawatzky}},\
  }\href {\doibase 10.1103/PhysRevB.41.288} {\bibfield  {journal} {\bibinfo
  {journal} {Phys. Rev. B}\ }\textbf {\bibinfo {volume} {41}},\ \bibinfo
  {pages} {288} (\bibinfo {year} {1990})}\BibitemShut {NoStop}%
\bibitem [{\citenamefont {Tjeng}\ \emph {et~al.}(1992)\citenamefont {Tjeng},
  \citenamefont {Chen},\ and\ \citenamefont {Cheong}}]{tjeng92a}%
  \BibitemOpen
  \bibfield  {author} {\bibinfo {author} {\bibfnamefont {L.~H.}\ \bibnamefont
  {Tjeng}}, \bibinfo {author} {\bibfnamefont {C.~T.}\ \bibnamefont {Chen}}, \
  and\ \bibinfo {author} {\bibfnamefont {S.-W.}\ \bibnamefont {Cheong}},\
  }\href {\doibase 10.1103/PhysRevB.45.8205} {\bibfield  {journal} {\bibinfo
  {journal} {Phys. Rev. B}\ }\textbf {\bibinfo {volume} {45}},\ \bibinfo
  {pages} {8205} (\bibinfo {year} {1992})}\BibitemShut {NoStop}%
\bibitem [{\citenamefont {Ebbinghaus}\ \emph {et~al.}(2009)\citenamefont
  {Ebbinghaus}, \citenamefont {Riegg}, \citenamefont {G\"otzfried},\ and\
  \citenamefont {Reller}}]{ebbinghaus10a}%
  \BibitemOpen
  \bibfield  {author} {\bibinfo {author} {\bibfnamefont {S.}~\bibnamefont
  {Ebbinghaus}}, \bibinfo {author} {\bibfnamefont {S.}~\bibnamefont {Riegg}},
  \bibinfo {author} {\bibfnamefont {T.}~\bibnamefont {G\"otzfried}}, \ and\
  \bibinfo {author} {\bibfnamefont {A.}~\bibnamefont {Reller}},\ }\href
  {\doibase 10.1140/epjst/e2010-01213-4} {\bibfield  {journal} {\bibinfo
  {journal} {Eur. Phys. J. Spec. Top.}\ }\textbf {\bibinfo {volume} {180}},\
  \bibinfo {pages} {91} (\bibinfo {year} {2009})}\BibitemShut {NoStop}%
\bibitem [{\citenamefont {Hu}\ \emph {et~al.}(2002)\citenamefont {Hu},
  \citenamefont {Golden}, \citenamefont {Ebbinghaus}, \citenamefont {Knupfer},
  \citenamefont {Fink}, \citenamefont {de~Groot},\ and\ \citenamefont
  {Kaindl}}]{hu02a}%
  \BibitemOpen
  \bibfield  {author} {\bibinfo {author} {\bibfnamefont {Z.}~\bibnamefont
  {Hu}}, \bibinfo {author} {\bibfnamefont {M.}~\bibnamefont {Golden}}, \bibinfo
  {author} {\bibfnamefont {S.}~\bibnamefont {Ebbinghaus}}, \bibinfo {author}
  {\bibfnamefont {M.}~\bibnamefont {Knupfer}}, \bibinfo {author} {\bibfnamefont
  {J.}~\bibnamefont {Fink}}, \bibinfo {author} {\bibfnamefont {F.}~\bibnamefont
  {de~Groot}}, \ and\ \bibinfo {author} {\bibfnamefont {G.}~\bibnamefont
  {Kaindl}},\ }\href {\doibase 10.1016/S0301-0104(02)00729-2} {\bibfield
  {journal} {\bibinfo  {journal} {Chem. Phys.}\ }\textbf {\bibinfo {volume}
  {282}},\ \bibinfo {pages} {451 } (\bibinfo {year} {2002})}\BibitemShut
  {NoStop}%
\bibitem [{\citenamefont {de~Groot}(1994)}]{degroot94a}%
  \BibitemOpen
  \bibfield  {author} {\bibinfo {author} {\bibfnamefont {F.}~\bibnamefont
  {de~Groot}},\ }\href {\doibase 10.1016/0368-2048(93)02041-J} {\bibfield
  {journal} {\bibinfo  {journal} {J. Electron Spectrosc. Relat. Phenom.}\
  }\textbf {\bibinfo {volume} {67}},\ \bibinfo {pages} {529 } (\bibinfo {year}
  {1994})}\BibitemShut {NoStop}%
\bibitem [{\citenamefont {Chang}\ \emph {et~al.}(2009)\citenamefont {Chang},
  \citenamefont {Hu}, \citenamefont {Wu}, \citenamefont {Burnus}, \citenamefont
  {Hollmann}, \citenamefont {Benomar}, \citenamefont {Lorenz}, \citenamefont
  {Tanaka}, \citenamefont {Lin}, \citenamefont {Hsieh}, \citenamefont {Chen},\
  and\ \citenamefont {Tjeng}}]{chang09a}%
  \BibitemOpen
  \bibfield  {author} {\bibinfo {author} {\bibfnamefont {C.~F.}\ \bibnamefont
  {Chang}}, \bibinfo {author} {\bibfnamefont {Z.}~\bibnamefont {Hu}}, \bibinfo
  {author} {\bibfnamefont {H.}~\bibnamefont {Wu}}, \bibinfo {author}
  {\bibfnamefont {T.}~\bibnamefont {Burnus}}, \bibinfo {author} {\bibfnamefont
  {N.}~\bibnamefont {Hollmann}}, \bibinfo {author} {\bibfnamefont
  {M.}~\bibnamefont {Benomar}}, \bibinfo {author} {\bibfnamefont
  {T.}~\bibnamefont {Lorenz}}, \bibinfo {author} {\bibfnamefont
  {A.}~\bibnamefont {Tanaka}}, \bibinfo {author} {\bibfnamefont {H.-J.}\
  \bibnamefont {Lin}}, \bibinfo {author} {\bibfnamefont {H.~H.}\ \bibnamefont
  {Hsieh}}, \bibinfo {author} {\bibfnamefont {C.~T.}\ \bibnamefont {Chen}}, \
  and\ \bibinfo {author} {\bibfnamefont {L.~H.}\ \bibnamefont {Tjeng}},\ }\href
  {\doibase 10.1103/PhysRevLett.102.116401} {\bibfield  {journal} {\bibinfo
  {journal} {Phys. Rev. Lett.}\ }\textbf {\bibinfo {volume} {102}},\ \bibinfo
  {pages} {116401} (\bibinfo {year} {2009})}\BibitemShut {NoStop}%
\bibitem [{\citenamefont {Schwingenschl\"ogl}\ \emph
  {et~al.}(2003)\citenamefont {Schwingenschl\"ogl}, \citenamefont {Eyert},\
  and\ \citenamefont {Eckern}}]{schwingenschloegl03a}%
  \BibitemOpen
  \bibfield  {author} {\bibinfo {author} {\bibfnamefont {U.}~\bibnamefont
  {Schwingenschl\"ogl}}, \bibinfo {author} {\bibfnamefont {V.}~\bibnamefont
  {Eyert}}, \ and\ \bibinfo {author} {\bibfnamefont {U.}~\bibnamefont
  {Eckern}},\ }\href {\doibase 10.1016/S0009-2614(03)00201-X} {\bibfield
  {journal} {\bibinfo  {journal} {Chem. Phys. Lett.}\ }\textbf {\bibinfo
  {volume} {370}},\ \bibinfo {pages} {719 } (\bibinfo {year}
  {2003})}\BibitemShut {NoStop}%
\bibitem [{\citenamefont {Yeh}\ and\ \citenamefont {Lindau}(1985)}]{yeh85a}%
  \BibitemOpen
  \bibfield  {author} {\bibinfo {author} {\bibfnamefont {J.}~\bibnamefont
  {Yeh}}\ and\ \bibinfo {author} {\bibfnamefont {I.}~\bibnamefont {Lindau}},\
  }\href {\doibase 10.1016/0092-640X(85)90016-6} {\bibfield  {journal}
  {\bibinfo  {journal} {At. Data Nucl. Data Tables}\ }\textbf {\bibinfo
  {volume} {32}},\ \bibinfo {pages} {1 } (\bibinfo {year} {1985})}\BibitemShut
  {NoStop}%
\bibitem [{\citenamefont {Ghijsen}\ \emph {et~al.}(1988)\citenamefont
  {Ghijsen}, \citenamefont {Tjeng}, \citenamefont {van Elp}, \citenamefont
  {Eskes}, \citenamefont {Westerink}, \citenamefont {Sawatzky},\ and\
  \citenamefont {Czyzyk}}]{ghijsen88a}%
  \BibitemOpen
  \bibfield  {author} {\bibinfo {author} {\bibfnamefont {J.}~\bibnamefont
  {Ghijsen}}, \bibinfo {author} {\bibfnamefont {L.~H.}\ \bibnamefont {Tjeng}},
  \bibinfo {author} {\bibfnamefont {J.}~\bibnamefont {van Elp}}, \bibinfo
  {author} {\bibfnamefont {H.}~\bibnamefont {Eskes}}, \bibinfo {author}
  {\bibfnamefont {J.}~\bibnamefont {Westerink}}, \bibinfo {author}
  {\bibfnamefont {G.~A.}\ \bibnamefont {Sawatzky}}, \ and\ \bibinfo {author}
  {\bibfnamefont {M.~T.}\ \bibnamefont {Czyzyk}},\ }\href {\doibase
  10.1103/PhysRevB.38.11322} {\bibfield  {journal} {\bibinfo  {journal} {Phys.
  Rev. B}\ }\textbf {\bibinfo {volume} {38}},\ \bibinfo {pages} {11322}
  (\bibinfo {year} {1988})}\BibitemShut {NoStop}%
\bibitem [{\citenamefont {Tjeng}\ \emph {et~al.}(1991)\citenamefont {Tjeng},
  \citenamefont {Chen}, \citenamefont {Ghijsen}, \citenamefont {Rudolf},\ and\
  \citenamefont {Sette}}]{tjeng91a}%
  \BibitemOpen
  \bibfield  {author} {\bibinfo {author} {\bibfnamefont {L.~H.}\ \bibnamefont
  {Tjeng}}, \bibinfo {author} {\bibfnamefont {C.~T.}\ \bibnamefont {Chen}},
  \bibinfo {author} {\bibfnamefont {J.}~\bibnamefont {Ghijsen}}, \bibinfo
  {author} {\bibfnamefont {P.}~\bibnamefont {Rudolf}}, \ and\ \bibinfo {author}
  {\bibfnamefont {F.}~\bibnamefont {Sette}},\ }\href {\doibase
  10.1103/PhysRevLett.67.501} {\bibfield  {journal} {\bibinfo  {journal} {Phys.
  Rev. Lett.}\ }\textbf {\bibinfo {volume} {67}},\ \bibinfo {pages} {501}
  (\bibinfo {year} {1991})}\BibitemShut {NoStop}%
\bibitem [{\citenamefont {Tjeng}\ \emph {et~al.}(1997)\citenamefont {Tjeng},
  \citenamefont {Sinkovic}, \citenamefont {Brookes}, \citenamefont {Goedkoop},
  \citenamefont {Hesper}, \citenamefont {Pellegrin}, \citenamefont {de~Groot},
  \citenamefont {Altieri}, \citenamefont {Hulbert}, \citenamefont {Shekel},\
  and\ \citenamefont {Sawatzky}}]{tjeng97a}%
  \BibitemOpen
  \bibfield  {author} {\bibinfo {author} {\bibfnamefont {L.~H.}\ \bibnamefont
  {Tjeng}}, \bibinfo {author} {\bibfnamefont {B.}~\bibnamefont {Sinkovic}},
  \bibinfo {author} {\bibfnamefont {N.~B.}\ \bibnamefont {Brookes}}, \bibinfo
  {author} {\bibfnamefont {J.~B.}\ \bibnamefont {Goedkoop}}, \bibinfo {author}
  {\bibfnamefont {R.}~\bibnamefont {Hesper}}, \bibinfo {author} {\bibfnamefont
  {E.}~\bibnamefont {Pellegrin}}, \bibinfo {author} {\bibfnamefont {F.~M.~F.}\
  \bibnamefont {de~Groot}}, \bibinfo {author} {\bibfnamefont {S.}~\bibnamefont
  {Altieri}}, \bibinfo {author} {\bibfnamefont {S.~L.}\ \bibnamefont
  {Hulbert}}, \bibinfo {author} {\bibfnamefont {E.}~\bibnamefont {Shekel}}, \
  and\ \bibinfo {author} {\bibfnamefont {G.~A.}\ \bibnamefont {Sawatzky}},\
  }\href {\doibase 10.1103/PhysRevLett.78.1126} {\bibfield  {journal} {\bibinfo
   {journal} {Phys. Rev. Lett.}\ }\textbf {\bibinfo {volume} {78}},\ \bibinfo
  {pages} {1126} (\bibinfo {year} {1997})}\BibitemShut {NoStop}%
\bibitem [{\citenamefont {Mizokawa}\ \emph {et~al.}(2009)\citenamefont
  {Mizokawa}, \citenamefont {Morita}, \citenamefont {Sudayama}, \citenamefont
  {Takubo}, \citenamefont {Yamada}, \citenamefont {Azuma}, \citenamefont
  {Takano},\ and\ \citenamefont {Shimakawa}}]{mizokawa09b}%
  \BibitemOpen
  \bibfield  {author} {\bibinfo {author} {\bibfnamefont {T.}~\bibnamefont
  {Mizokawa}}, \bibinfo {author} {\bibfnamefont {Y.}~\bibnamefont {Morita}},
  \bibinfo {author} {\bibfnamefont {T.}~\bibnamefont {Sudayama}}, \bibinfo
  {author} {\bibfnamefont {K.}~\bibnamefont {Takubo}}, \bibinfo {author}
  {\bibfnamefont {I.}~\bibnamefont {Yamada}}, \bibinfo {author} {\bibfnamefont
  {M.}~\bibnamefont {Azuma}}, \bibinfo {author} {\bibfnamefont
  {M.}~\bibnamefont {Takano}}, \ and\ \bibinfo {author} {\bibfnamefont
  {Y.}~\bibnamefont {Shimakawa}},\ }\href {\doibase 10.1103/PhysRevB.80.125105}
  {\bibfield  {journal} {\bibinfo  {journal} {Phys. Rev. B}\ }\textbf {\bibinfo
  {volume} {80}},\ \bibinfo {pages} {125105} (\bibinfo {year}
  {2009})}\BibitemShut {NoStop}%
\bibitem [{\citenamefont {Long}\ \emph
  {et~al.}(2009{\natexlab{a}})\citenamefont {Long}, \citenamefont {Saito},
  \citenamefont {Mizumaki}, \citenamefont {Agui},\ and\ \citenamefont
  {Shimakawa}}]{long09b}%
  \BibitemOpen
  \bibfield  {author} {\bibinfo {author} {\bibfnamefont {Y.}~\bibnamefont
  {Long}}, \bibinfo {author} {\bibfnamefont {T.}~\bibnamefont {Saito}},
  \bibinfo {author} {\bibfnamefont {M.}~\bibnamefont {Mizumaki}}, \bibinfo
  {author} {\bibfnamefont {A.}~\bibnamefont {Agui}}, \ and\ \bibinfo {author}
  {\bibfnamefont {Y.}~\bibnamefont {Shimakawa}},\ }\href {\doibase
  10.1021/ja906668c} {\bibfield  {journal} {\bibinfo  {journal} {J. Am. Chem.
  Soc.}\ }\textbf {\bibinfo {volume} {131}},\ \bibinfo {pages} {16244}
  (\bibinfo {year} {2009}{\natexlab{a}})}\BibitemShut {NoStop}%
\bibitem [{\citenamefont {Morita}\ \emph {et~al.}(2010)\citenamefont {Morita},
  \citenamefont {Sudayama}, \citenamefont {Takubo}, \citenamefont {Shiraki},
  \citenamefont {Saito}, \citenamefont {Shimakawa},\ and\ \citenamefont
  {Mizokawa}}]{morita10a}%
  \BibitemOpen
  \bibfield  {author} {\bibinfo {author} {\bibfnamefont {Y.}~\bibnamefont
  {Morita}}, \bibinfo {author} {\bibfnamefont {T.}~\bibnamefont {Sudayama}},
  \bibinfo {author} {\bibfnamefont {K.}~\bibnamefont {Takubo}}, \bibinfo
  {author} {\bibfnamefont {H.}~\bibnamefont {Shiraki}}, \bibinfo {author}
  {\bibfnamefont {T.}~\bibnamefont {Saito}}, \bibinfo {author} {\bibfnamefont
  {Y.}~\bibnamefont {Shimakawa}}, \ and\ \bibinfo {author} {\bibfnamefont
  {T.}~\bibnamefont {Mizokawa}},\ }\href {\doibase 10.1103/PhysRevB.81.165111}
  {\bibfield  {journal} {\bibinfo  {journal} {Phys. Rev. B}\ }\textbf {\bibinfo
  {volume} {81}},\ \bibinfo {pages} {165111} (\bibinfo {year}
  {2010})}\BibitemShut {NoStop}%
\bibitem [{\citenamefont {Long}\ \emph
  {et~al.}(2009{\natexlab{b}})\citenamefont {Long}, \citenamefont {Hayashi},
  \citenamefont {Saito}, \citenamefont {Azuma}, \citenamefont {Muranaka},\ and\
  \citenamefont {Shimakawa}}]{long09a}%
  \BibitemOpen
  \bibfield  {author} {\bibinfo {author} {\bibfnamefont {Y.~W.}\ \bibnamefont
  {Long}}, \bibinfo {author} {\bibfnamefont {N.}~\bibnamefont {Hayashi}},
  \bibinfo {author} {\bibfnamefont {T.}~\bibnamefont {Saito}}, \bibinfo
  {author} {\bibfnamefont {M.}~\bibnamefont {Azuma}}, \bibinfo {author}
  {\bibfnamefont {S.}~\bibnamefont {Muranaka}}, \ and\ \bibinfo {author}
  {\bibfnamefont {Y.}~\bibnamefont {Shimakawa}},\ }\href@noop {} {\bibfield
  {journal} {\bibinfo  {journal} {Nature}\ }\textbf {\bibinfo {volume} {458}},\
  \bibinfo {pages} {60} (\bibinfo {year} {2009}{\natexlab{b}})}\BibitemShut
  {NoStop}%
\bibitem [{\citenamefont {Shimakawa}\ and\ \citenamefont
  {Takano}(2009)}]{shimakawa09a}%
  \BibitemOpen
  \bibfield  {author} {\bibinfo {author} {\bibfnamefont {Y.}~\bibnamefont
  {Shimakawa}}\ and\ \bibinfo {author} {\bibfnamefont {M.}~\bibnamefont
  {Takano}},\ }\href {\doibase 10.1002/zaac.200900248} {\bibfield  {journal}
  {\bibinfo  {journal} {Z. Anorg. Allg. Chem.}\ }\textbf {\bibinfo {volume}
  {635}},\ \bibinfo {pages} {1882} (\bibinfo {year} {2009})}\BibitemShut
  {NoStop}%
\bibitem [{\citenamefont {Ramirez}\ \emph {et~al.}(2004)\citenamefont
  {Ramirez}, \citenamefont {Lawes}, \citenamefont {Li},\ and\ \citenamefont
  {Subramanian}}]{ramirez04a}%
  \BibitemOpen
  \bibfield  {author} {\bibinfo {author} {\bibfnamefont {A.}~\bibnamefont
  {Ramirez}}, \bibinfo {author} {\bibfnamefont {G.}~\bibnamefont {Lawes}},
  \bibinfo {author} {\bibfnamefont {D.}~\bibnamefont {Li}}, \ and\ \bibinfo
  {author} {\bibfnamefont {M.}~\bibnamefont {Subramanian}},\ }\href {\doibase
  10.1016/j.ssc.2004.04.014} {\bibfield  {journal} {\bibinfo  {journal} {Solid
  State Commun.}\ }\textbf {\bibinfo {volume} {131}},\ \bibinfo {pages} {251 }
  (\bibinfo {year} {2004})}\BibitemShut {NoStop}%
\bibitem [{\citenamefont {Terasaki}\ \emph {et~al.}(2010)\citenamefont
  {Terasaki}, \citenamefont {Iwakawa}, \citenamefont {Nakano}, \citenamefont
  {Tsukuda},\ and\ \citenamefont {Kobayashi}}]{terasaki10a}%
  \BibitemOpen
  \bibfield  {author} {\bibinfo {author} {\bibfnamefont {I.}~\bibnamefont
  {Terasaki}}, \bibinfo {author} {\bibfnamefont {M.}~\bibnamefont {Iwakawa}},
  \bibinfo {author} {\bibfnamefont {T.}~\bibnamefont {Nakano}}, \bibinfo
  {author} {\bibfnamefont {A.}~\bibnamefont {Tsukuda}}, \ and\ \bibinfo
  {author} {\bibfnamefont {W.}~\bibnamefont {Kobayashi}},\ }\href {\doibase
  10.1039/B914661J} {\bibfield  {journal} {\bibinfo  {journal} {Dalton Trans.}\
  }\textbf {\bibinfo {volume} {39}},\  (\bibinfo {year} {2010})}\BibitemShut
  {NoStop}%
\bibitem [{\citenamefont {Zhang}\ and\ \citenamefont {Rice}(1988)}]{zhang88a}%
  \BibitemOpen
  \bibfield  {author} {\bibinfo {author} {\bibfnamefont {F.~C.}\ \bibnamefont
  {Zhang}}\ and\ \bibinfo {author} {\bibfnamefont {T.~M.}\ \bibnamefont
  {Rice}},\ }\href {\doibase 10.1103/PhysRevB.37.3759} {\bibfield  {journal}
  {\bibinfo  {journal} {Phys. Rev. B}\ }\textbf {\bibinfo {volume} {37}},\
  \bibinfo {pages} {3759} (\bibinfo {year} {1988})}\BibitemShut {NoStop}%
\bibitem [{\citenamefont {Eskes}\ and\ \citenamefont
  {Sawatzky}(1988)}]{eskes88a}%
  \BibitemOpen
  \bibfield  {author} {\bibinfo {author} {\bibfnamefont {H.}~\bibnamefont
  {Eskes}}\ and\ \bibinfo {author} {\bibfnamefont {G.~A.}\ \bibnamefont
  {Sawatzky}},\ }\href {\doibase 10.1103/PhysRevLett.61.1415} {\bibfield
  {journal} {\bibinfo  {journal} {Phys. Rev. Lett.}\ }\textbf {\bibinfo
  {volume} {61}},\ \bibinfo {pages} {1415} (\bibinfo {year}
  {1988})}\BibitemShut {NoStop}%
\bibitem [{\citenamefont {Brookes}\ \emph {et~al.}(2001)\citenamefont
  {Brookes}, \citenamefont {Ghiringhelli}, \citenamefont {Tjernberg},
  \citenamefont {Tjeng}, \citenamefont {Mizokawa}, \citenamefont {Li},\ and\
  \citenamefont {Menovsky}}]{brookes01a}%
  \BibitemOpen
  \bibfield  {author} {\bibinfo {author} {\bibfnamefont {N.~B.}\ \bibnamefont
  {Brookes}}, \bibinfo {author} {\bibfnamefont {G.}~\bibnamefont
  {Ghiringhelli}}, \bibinfo {author} {\bibfnamefont {O.}~\bibnamefont
  {Tjernberg}}, \bibinfo {author} {\bibfnamefont {L.~H.}\ \bibnamefont
  {Tjeng}}, \bibinfo {author} {\bibfnamefont {T.}~\bibnamefont {Mizokawa}},
  \bibinfo {author} {\bibfnamefont {T.~W.}\ \bibnamefont {Li}}, \ and\ \bibinfo
  {author} {\bibfnamefont {A.~A.}\ \bibnamefont {Menovsky}},\ }\href {\doibase
  10.1103/PhysRevLett.87.237003} {\bibfield  {journal} {\bibinfo  {journal}
  {Phys. Rev. Lett.}\ }\textbf {\bibinfo {volume} {87}},\ \bibinfo {pages}
  {237003} (\bibinfo {year} {2001})}\BibitemShut {NoStop}%
\end{thebibliography}
\end{document}